\documentclass[12pt]{article}
\usepackage{epsfig}

\parskip=\medskipamount    
\renewcommand{\thesection}{\arabic{section}}

\catcode`@=11

\@addtoreset{equation}{section}
\@addtoreset{equation}{subsection}
\def\theequation{\ifnum\value{section}=0 \arabic{equation}\ignorespaces
\else \ifnum\value{section}=-1 A.\arabic{equation}\ignorespaces
\else \ifnum\value{subsection}=0 \thesection.\arabic{equation}\ignorespaces
\else \thesection.\arabic{subsection}.\arabic{equation}\ignorespaces
              \fi
                        \fi
                   \fi}

{\catcode`\'=\active \def'{{}^\bgroup\prim@s}}

\catcode`@=12

\newcommand{\bq}{\begin{equation}}
\newcommand{\be}{\begin{equation}} 
\newcommand{\fq}{\end{equation}}
\newcommand{\ee}{\end{equation}}
\newcommand{\bqr}{\begin{eqnarray}}
\newcommand{\beqs}{\begin{eqnarray}} 
\newcommand{\fqr}{\end{eqnarray}}
\newcommand{\eeqs}{\end{eqnarray}}

\newcommand{\rf}[1]{(\ref{#1})}

\def\pa{\partial}

\def\bop#1{\setbox0=\hbox{$#1M$}\mkern1.5mu
  \vbox{\hrule height0pt depth.04\ht0
    \hbox{\vrule width.04\ht0 height.9\ht0 \kern.9\ht0
    \vrule width.04\ht0}\hrule height.04\ht0}\mkern1.5mu}
\def\Box{{\mathpalette\bop{}}}                    

\def\Prod{\prod}

\begin{document}
\thispagestyle{empty} 

\begin{flushright}
\begin{tabular}{l} 
UCLA-02-TEP-27\\
hep-th/0209086 \\ 
\end{tabular} 
\end{flushright}  

\vskip .6in 
\begin{center} 

{\Large\bf Gauge Theories in the Derivative Expansion} 

\vskip .5in 

{\bf Gordon Chalmers} 
\\[5mm] 
{\em Department of Physics and Astronomy \\ 
University of California at Los Angeles \\ 
Los Angeles, CA  90025-1547 } \\  

{e-mail: chalmers@physics.ucla.edu}  

\vskip .5in minus .2in 

{\bf Abstract}   
\end{center} 

Gauge theories with and without matter are formulated in the derivative 
expansion.  Amplitudes are derived as a power series in the energy 
scales; there are simplifications as compared with the usual loop 
expansion.  The incorporation and summation of Wilson loops is natural 
in this approach and allows for a derivation of the confining potential as 
well as the masses of the gauge invariant observables.

\setcounter{page}{0}
\newpage 
\setcounter{footnote}{0} 

\section{Introduction} 

In this work we derive, in the derivative expansion, the correlation 
functions and amplitudes of conventional QCD.  The masses of the 
mesonic, hadronic and further states together with their dynamics are 
obtainable through this approach.  In subsequent work we analyze 
a holographic version, incorporating gravity, although conventional 
quantum chromodynamics meshes with conventional gravity.  

The composite operators of gauge theory, including matter, are, 

\bqr 
\Prod_{i=1}^n \Prod_{j=1}^{n^\partial_F} \nabla_{\mu_{sigma}} 
\Prod_{j=1}^{n^F} F_{\mu_{\sigma(i,j)},\nu_{\sigma'(i,j)}}
\Prod_{i=1}^{n^\partial_\phi} \nabla_{\mu_{\sigma(i)}} \Prod_{j=1}^{n_\phi} 
\Phi 
\fqr 
\bqr \times 
\Prod_{i=1}^{n^\partial_q} \nabla_{\mu_{\sigma(i)}}
  \Prod_{j=1}^{n_\psi} q_{\alpha_{\sigma(i,j)}} \ .
\fqr 
We focus on the composite operators, ${\cal O}_{m}$, of dimension $n$, 
that are pure gluonic (glueballs) or quark-like.  

The amplitudes may be constructed via two methods.  The conventional 
methodology is to sum over the Feynman diagrams.  Recently, the 
derivative expansion has been utilized to obtain the same information 
in scalar and spontaneously broken $N=4$ gauge theory; we take 
the second approach in this work, together with non-perturbative 
corrections, in analyzing quantum chromodynaimcs.  The derivative 
expansion has the advantage 
that the restricted set of diagrams that enter may all be integrated 
over, and that the non-perturbative corrections are easily implemented.  

The masses and confining potentials of the nucleons may be found.  
Furthermore, the correlations of the
quantum dressed composite operators are found.  The methodology is the 
same as the one used in analyzing (and almost solving) the $\phi^n$ 
theory, but with the complication that the integrals have massless 
particles.  

The Langrangian we consider is 

\bqr 
{\cal L} = \int d^4x \bigl(-{1\over 4} F^{a}F_{a} + 
  \psi^{\dagger,a}\gamma^u\nabla_u\psi_a + \sum m_j \bar\psi_j \psi_j  \bigr) 
\ . 
\fqr 
In the case of QCD we have the usual doublet-triplets (t, b, c) and 
(s, d, u).  The masses are, in terms of $\Lambda (\Lambda/m_{pl})^{n_j/16}$, 
with $\Lambda=2$ TeV, 

\bqr 
\pmatrix{ 
1 & m_t \cr 
2 & m_b \cr 
3 & m_c \cr 
4 & m_s \cr 
5 & m_d \cr 
6 & m_u } 
\fqr   
and likewise for the leptons, 

\bqr 
\pmatrix{ 
3 & m_\tau \cr 
4 & m_\mu \cr 
5 & m_e } \ .
\fqr 
The quantization of the masses follows from gravitational considerations 
\cite{Chalmersgrav}.

The derivative expansion follows the simpler analysis of scalar field 
theory \cite{Chalmersscalar}. First we categorize the quantum generating 
functional of the 
S-matrix.  In the case of gauge theory, the quantum vertices are 
non-polynomial due to zero-mass thresholds, and inverse powers of 
derivatives are present due to massless singularities.  These two 
features, together with index structure, complicate the sewing that 
generates the S-matrix.  However, the coefficients between different 
orders are not all independent due to known factorization identities.

The quantum generating functional is found from the general vertex 
of hard dimension, 

\bqr 
\prod_{i=1}^{p^A} \prod_{j=1}^{n_i^A} \pa_{\mu_{\sigma(i,j)}} 
  \prod^{m_i^A} A_{\mu_{\tilde\sigma(i,j)}}
\prod_{i=1}^{p^\psi} \prod_{j=1}^{n_i^\psi} \pa_{\mu_{\sigma(i,j)}}
  \prod^{m_i^\psi} \psi_{\alpha_{\tilde\sigma(i,j)}}  
\label{vertices}
\fqr 
with $\sum m_i^A=n^A$- and $\sum m_i^\psi =n^\psi$-point.  The 
effective action is gauge invariant but gauge dependent (e.g. 
$\lambda {\rm Tr} F^4$) off shell, and the soft dimensional modifications are 

generic.  The latter model the corrections to the classical scaling 
dimensions and have the form, 

\bqr 
\ln^m(\Box), \ln\ln\ldots\ln(\Box) \ , 
\fqr 
together with combinations.  They are placed generically within the hard 
dimension vertex.  Last, due to infra-red singularities (the form is 
known at multi-loop), we may place a $1/\Box$ in front 
of $A^p$ with $p$ arbitrary.  Massive particles have neither in the 
derivative expansion at low energy.

The polylogarithms encountered in multi-loop gauge theory 
calculations may be expanded upon this set of functions.  Calculation 
of the integrals that contain these terms is facilitated by a useful 
analytic continuation \cite{Chalmers:1997gn,Bender:rq},  

\bqr 
\Box^{\beta} = \int_0^\infty dt t^{-1+\beta} e^{-t\Box} \ , 
\fqr 
in which the exponential acts on the internal lines. The logarithms 
are found then by expanding, 

\bqr 
\Box^{\beta} = 1 +\beta \ln(\Box^2) + {1\over 2} \beta^2 \ln^2(\Box) 
   + \ldots \ . 
\fqr 
The nested log terms are handled via, 

\bqr 
(\Box^{{\beta}_1})^{\beta_2} = 1 + \beta_2\ln^{\beta_1}(\Box) + \ldots 
  = 1 + \beta_2 (1+\beta_1 \ln\ln(\Box) + \ldots) + \ldots \ ,
\fqr 
with calculations done at integral values and then continued to 
obtain the logarithms.  This analyticity has been usefule in studying 
scalar field theory and quantum electrodynamics.   The general 
effective theory is built from the complete set of polynomials and 
these two modifications.  Furthermore, the color structure is generically 
multi-trace in these vertices, although a truncation to leading color 
(planar) 
is consistent. 

With these sets of vertices, we may derive the recursion formulae to 
obtain the coefficients.  In exchange for complicated multi-loop Feynman 
diagrams, which have never been computed at three-loops,  there is  
complicated tensorial algebra; the integrals are all computable, however, 
and the algebra is suited for computer calculations.  

The gauge boson sewing and recursion formulae have the form, at 
$m_l+m_r=n$-point,  

\bqr 
\sum_L  \left[ \Prod_{i=1}^{m_l} \left( \Prod_{j=1}^{m_i^\pa} 
    \pa_{\mu_{\sigma(i,j)}} A_{\mu_i}(k_i) \right) \right]  
\left[ \Prod_{i=m_l+1}^{m_r} \left( \Prod_{j=1}^{{\tilde m}_i^\pa}
       \pa_{\mu_{{\tilde\sigma}(i,j)}} A_{\nu_i}(k_i) \right) \right]
\fqr 
\bqr  
\times 
\langle \left[ \Prod_{i=1}^L \left( \Prod_{j=1}^{m_i^\pa}
        \pa_{\mu_{\sigma'(i,j)}} \Prod_{j=1}^{m_i^A} A_{\mu_i}
        \right) \right] 
\fqr
\bqr 
\times 
    \left[ \Prod_{i=1}^L \left( \Prod_{j=1}^{{\tilde m}_i^\pa}
  \pa_{\mu_{{\tilde\sigma}'(i,j)}} \Prod_{j=1}^{{\tilde m}_i^A}
        A_{\mu_i} \right) \right] 
\rangle \times   t^{\mu_\sigma,\mu_i,m_i^\pa} 
t^{\tilde\mu_\sigma,{\tilde\mu}_i,{\tilde m}_i^\pa}  + {\rm perms} \ , 
\label{loop}
\fqr 
with $\sum m_i^A=\sum {\tilde m}_i^A=L$.  The coupling dependence 
in eqn. \rf{loop} is within the tensor $t$, and is a power series, 

\bqr 
t^{\mu_\sigma,\mu_i,m_i^\pa} = \sum_{m=n-3}^\infty 
t^{\mu_\sigma,\mu_i,m_i^\pa}_{m} g_{YM}^{2m} \ ,  
\fqr 
at $n$-point.  The logarithmic modifications enter via the functions 
$f_i(\Box)$ placed within the expression.  The contraction of the 
internal lines is straightforward.   The recursion formulae is derived 
by equating eqns. \rf{loop} with \rf{vertices}.  

\section{Non-perturbative}

In the previous section the coupling of microscopic gauge fields and 
fermions was examined in perturbation theory, and the beta function 
controls the scaling of the coupling constant.  

We examine more general configurations in this section and in 
particular, the contribution of the analog of the Wilson loop to the 
amplitudes.  Summing these configurations generates a confining 
potential; these coherent loops generate the reggeization 
of the composite states through the Schroedinger equation, modeled 
by the composite operators.  The resonances of 
the mesons, baryons, glueballs, and the like, are also potentially 
obtained.     

The general correlation of the microscopic theory is 

\bqr 
\langle \Prod_{i=1}^n A^{a_i}_{\mu_i} 
  \Prod_{j=1}^m e^{-{\alpha^{1/2}\over g} \oint A} \rangle \ , 
\fqr 
where the multiple traces of the exponentials are implied.  
The coupling $g$ appears in the coherent state, and $\alpha$ 
is a dimensionful parameter chosen so that the calculated 
pion ($u{\bar d}-{\bar u}d$) mass has the expected value.  
In asymptotically free theories the coupling $g$ runs to zero; 
for simplicity we shall absorb it into $\alpha$.  Composite 
operator correlators are expectations, 

\bqr 
\langle \Prod_{i=1}^n {\cal O}_{(d_i)}^{(i)} \Prod_{j=1}^m  
    e^{-{\alpha^{1/2}\over g}\oint A} \rangle 
\fqr 
The diagrams and integrals representing the perturbative calculations 
of the microscopic and gauge invariant correlatorion are different, 
although both resemble free-field theory.  The correlation functions 
are found from the effective theory by sewing the composite operator's 
external legs to other composite operators, without iteration.  The 
full two-point function is to be utilized for this purpose.  In general 
the soft dimensions of the operators that model the non-classical 
scaling dimensions are included.   

First lets examine the closed analog of the Wilson lines, as they have 
a dramatic effect on the amplitudes.  The expression 

\bqr 
e^{-{\alpha^{1/2}\over g}\oint A} e^{-{\alpha^{1/2}\over g}\oint A} 
\fqr 
integrates, or contracts, to 

\bqr 
e^{{m\beta_L (L+1) g^2}{ \vert x\vert \over\alpha x^2}} = 
  e^{{m\beta_L (L+1)} {g^2 \over \alpha\vert x\vert}} \ .  
\label{contract} 
\fqr 
The parameter $\beta$ is a number representing a constant factor 
obtained  from the integraction eqn. \rf{contract}.   The form in eqn. 
\rf{contract}  is similar to a mass term, with mass $\alpha/beta$.  
The diagram is illustrated in fig. 1; in general, these closed loops 
intersect an arbitrary number $m$ times from $2$ to $\infty$.  These 
multiple points are to be integrated over, resulting in a factor of 
$\vert x\vert$, as the perturbative graphs are of "thickness" 
$x_1-x_2=x$.  These Wilson loops receive corrections via 
perturbation theory between the lines.  Multiple non-interacting 
$e^{-{\alpha^{1/2}\over g} \oint A}$ loops, as represented in figure 2, 
generate 

\bqr 
e^{{n (L+1)}{\beta_L\over\alpha\vert x\vert}} \ .  
\label{closed} 
\fqr 
A mass term is $e^{-mx}$, which does compare with the $e^{1/x^2}$.  
The summation over intersections between the Wilson line and the 
loop graph is straightforward:  The loop has three choices upon 
encountering the graph, under/over or intersection.  This results in a 
combinatoric factor of $3^p$ for the result in \rf{closed}, when 
there are $p$ intersections.  Count the number of loops via the 
intersection number: one example is a product of minimal loops 
(a of them) and another is one loop with 2a intersections.  The expected 
length of the loop is the number of intersections divided by two, and 
times the distance $x_1\cdot x_2$ in the diagram; this is a rough 
approximation, and we have neglected the interactions between the 
closed loops.  

This formulae in \rf{closed} suggests a weakening of the masses; however, we 
have an infinite summation of these closed loops, together with interactions 
between.  The individual term in eqn. \rf{closed} when expanded at small $x$, 

gives a $1/x$ term at small separation.  The exponentials will be resummed, 
however, and the short distance singular Couolomb interaction comes from 
the perturbative results $\langle \psi\psi\ldots\psi$.   

At loop $L$, there is a symmetry factor of $3^{2L}/L!$ that would 
effect the $L$ structure.  The sum over the closed non-interacting 
loops is, 

\bqr 
\sum_{n=0}^\infty e^{{n(L+1)}{\beta\over \alpha x}} 
  = {1\over 1 - e^{\beta\over\alpha x}} \ .  
\fqr 
This potential at small and large $x^2$ is, 

\bqr 
V(x)= e^{-{\beta(L+1)}{ g^2\over\alpha x}} + {\cal O}(e^{-2/x}) \ , 
\fqr 
\bqr  
V(x)= {\beta (L+1)}{ x g^2\over\alpha}+{\cal O}(x^4) \ .
\fqr 
These results modify the perturbative calculations.  Different 
loop orders have different kinematic due to the integrations; 
however, as the $L$ dependence goes like $L/2^L$ due to combinatorics 
the large $x$ limit should not change much except for a factor.  

The expected confining potential of the quarks and gluons is
generated.  At weak coupling, the exponential may be expanded 
in a power series.  The first term is from tree-level perturbation theory 
($e^\alpha=1$) modeling the Coulomb interaction, 

\bqr 
V(x) ={g^2\over {\sqrt\alpha} x} + {{\sqrt\alpha} \over g}  
  + \ldots  \ . 
\fqr 
The constant factor is dropped, and in the general coupling case, we 
keep the $e^{-1/x}$.  As a result of the potential, the composite 
operators have a well defined meaning in terms of partonic 
constituents.

There are modifications of the above result due to interactions 
between the closed lines, illustrated in figure (3).  Via the confining 
potential, we find the excited (not lowest spin) states 
associated with binding $m$ quarks.  Of course, there are 
modifications of the potential due to the perturbative loops, 
both in the derivative expansion diagrams and in the 
exponential.

In the scattering functions we integrate over $x$.  The closed 
exponentiated loops are integrable at $x^2\rightarrow 0$ due to 
the $e^{-1/\vert x\vert}$.  This factor also modifies the high energy 
behavior by removing the ultra-violet divergences within the integrals.

The running of the coupling constant follows as in perturbation theory, 
including theories with different matter content.  In the ultra-violet,
if the coupling runs to zero then the Wilson loops become exponentially 
suppressed and break apart.  However, the resummed series smooths out 
the ultra-violet divergences at intermediate couplings as a string would 
do. 

The amplitudes and quark/gluon potentials depend on the details 
of the microscopic theory and the couplings within it.  

\section{Discussion} 

The derivative expansion in the context of gauge theories with 
general matter content is studied.  The expansion is developed 
and recursion formulae are developed in order to determine the
coefficients of the expansion.  The integrals are much simpler 
than in the usual perturbative approach, in line with recent 
results pertaining to scalar and spontaneously broken 
supersymmetric gauge theories.  

The interactions between the partonic constituents are examined 
upon including the analog of Wilson loops in this approach.  The 
potential derived after summing the latter loops generates a 
confining potential and a reggeization of the composite states, 
including the glueballs and quark matter.   

\section*{Acknowledgements} 

GC thanks the DOD, 444025-HL-25619, for support.

\end{document}